\documentclass[9pt]{article}%
\usepackage[conf]{ametsoc}
\usepackage{graphicx}
\usepackage{amssymb}
\usepackage{amsmath}
\usepackage{amsfonts}
\usepackage{hyperref}%
\providecommand{\U}[1]{\protect\rule{.1in}{.1in}}
\renewcommand{\cite}[2][]{\citep[#1]{#2}}
\begin{document}

\title{WRF fire simulation coupled with a fuel moisture model and smoke transport by
WRF-Chem\footnotemark[1]}
\author{Adam K. Kochanski\footnotemark[2], Jonathan D. Beezley\footnotemark[3], Jan
Mandel\footnotemark[3], and Minjeong Kim\footnotemark[3]}
\maketitle

\footnotetext[1]{WRF Users Workshop 2012. This research was supported by the
National Science Foundation under grant AGS-0835579, and by U.S. National
Institute of Standards and Technology Fire Research Grants Program grant
60NANB7D6144.}

\footnotetext[2]{Department of Meteorology, University of Utah, Salt Lake
City, UT}

\footnotetext[3]{Department of Mathematical and Statistical Sciences,
University of Colorado Denver, Denver, CO}

\section{Introduction}

We describe two recent additions to SFIRE, a fire spread model coupled with
WRF \cite{Mandel-2009-DAW,Mandel-2011-CAF}. This model builds on the earlier
CAWFE code \cite{Clark-1996-CAF-x,Clark-2004-DCA,Coen-2005-SBE}; see
\cite{Mandel-2011-CAF} and
\url{http://www.openwfm.org/wiki/WRF-Fire_development_notes} for further
historical details and acknowledgements. The coupled model is available from
\url{OpenWFM.org}. An earlier version of the code is included in WRF release
as WRF-Fire.


\section{Fuel moisture model}

\label{sec:moisture}

Fire spread rate depends strongly on the moisture contents of the fuel. In
fact, the spread rate drops to zero when the moisture reaches the so-called
extinction value \cite{Pyne-1996-IWF}. For this reason, we have coupled the
fire spread model with a simple fuel moisture model integrated in SFIRE and
run independently at points of the mesh. See
\cite{Nelson-2000-PDC,Weise-2005-CTM} for other, much more sophisticated
models. \cite{Coen-2005-SBE} used an assumed diurnal dependence of fuel moisture
on time. Empirical models
\cite{Fosberg-1971-DHT,VanWagner-1985-EFP} attempt to predict fuel moisture
from meteorological conditions measured daily. We use a simple timelag
differential equation at every point of the domain. This equations has
solutions which approach an equilibrium exponentially, if the equilibrium does
not change. In general, the solutions track a changing equilibrium with a delay.

\label{sec:equilibrium}First, relative humidity is computed from the
atmospheric temperature $T$ (K), waver vapor contents $Q$ (kg$/$kg), and
pressure $P$ (Pa). The saturated water vapor pressure $P_{WS}$ (Pa) is
approximated, following \cite[Eq. (10)]{Murphy-2005-RVP}, as $P_{WS}%
=\exp(54.842763-6763.22/T-4.210\log T+0.000367T+\tanh\left\{  0.0415\left(
T-218.8\right)  \right\}  (53.878-1331.22/T-9.44523\log
T+0.014025T)),123<T<332\mathrm{K}$. The water vapor pressure is $P_{W}=PQ/\left(
\varepsilon+\left(  \ 1-\varepsilon\right) Q \right)  $, where $\varepsilon
=0.622$ is the ratio of the molecular weight of water ($18.02$ g/mol) to
molecular weight of dry air ($28.97$ g/mol). We then obtain the relative
humidity $H=100P_{W}/P_{WS}.$The temperature and the relative humidity of the
air are then used to estimate the drying and wetting fuel equilibrium moisture
contents \cite[eq. (4), (5)]{VanWagner-1985-EFP} \cite[eq. (7), (8)]%
{Viney-1991-RFF},
\begin{align*}
E_{d}  &  =0.924H^{0.679}+0.000499e^{0.1H}+0.18(21.1+273.15-T)(1-e^{-0.115H}%
),\\
E_{w}  &  =0.618H^{0.753}+0.000454e^{0.1H}+0.18(21.1+273.15-T)(1-e^{-0.115H}).
\end{align*}

The fuel is considered as a combination of time-lag classes, and the fuel
moisture contents $m_{k}\left(  t\right)  $ in each class $k=1,\ldots,N$ is
then modeled by the standard time-lag equation%
\begin{equation}
\frac{dm_{k}}{dt}=\left\{
\begin{array}
[c]{c}%
\frac{E_{d}-m_{k}}{T_{k}}\text{ if }m_{k}>E_{d}\\
0\text{ if }E_{d}<m_{k}<E_{w}\\
\frac{E_{w}-m_{k}}{T_{k}}\text{ if }m_{k}<E_{w}%
\end{array}
\right.  \label{eq:moisture-ode}%
\end{equation}
where $T_{k}$ is the lag time. We use the standard model with the fuel
consisting of components with $T_{k}=1$, $10$, and $100$ hour lag time, with
the proportions $w_{k}\geq0,$ $\sum_{k=1}^{N}w_{k}=1$, given by the fuel
category description \cite{Scott-2005-SFB}. The overall fuel moisture then is
the weighted average $m=\sum_{k=1}^{N}w_{k}m_{k}$.

\begin{figure}[ptb]
\begin{center}%
\begin{tabular}
[c]{cc}%
\includegraphics[width=3in]{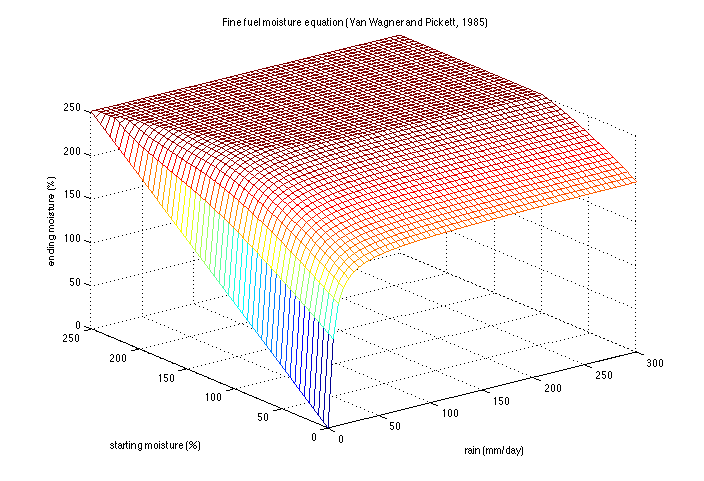} &
\includegraphics[width=3in]{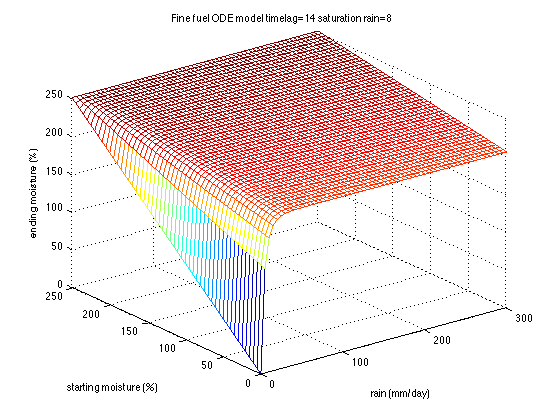}\\
(a) & (b)
\end{tabular}
\end{center}
\caption{Response of fine fuels to rain over 24 hours (a) following
\cite{VanWagner-1985-EFP} (b) from the time-lag model (\ref{eq:rain-ode}) by a
calibration of coefficients.}%
\label{fig:finerain}%
\end{figure}

During rain, the equilibrium moisture $E_{d}$ or $E_{w}$ is replaced by the
saturation moisture contents $S$, and equation (\ref{eq:moisture-ode}) is
modified to achieve the rain-wetting lag time $T_{rk}$ for heavy rain only
asymptotically, when the rain intensity $r$ (mm/h) is large:%
\begin{equation}
\frac{dm_{k}}{dt}=\frac{S-m_{k}}{T_{rk}}\left(  1-\exp\left(  -\frac{r-r_{0}%
}{r_{k}}\right)  \right)  ,\text{ if }r>r_{0}, \label{eq:rain-ode}%
\end{equation}
where $r_{0}$ is the threshold rain intensity below which no perceptible
wetting occurs, and $r_{k}$ is the saturation rain intensity for fuel
component $k$. At the saturation rain intensity, $1-1/e\approx63\%$ of the
maximal rain-wetting rate is achieved. The coefficients can be calibrated to
achieve a similar behavior as accepted empirical models
\cite{Fosberg-1971-DHT,VanWagner-1985-EFP}. This model estimates the fuel
moisture as a function of the initial moisture contents and rain accumulation
over 24 hours. Assuming steady rain over the 24 hours, we have obtained a
reasonable match with the Canadian fire danger rating system
\cite{VanWagner-1985-EFP} using $S=250\%,T_{rk}=14$ h, $r_{0}=0.05$ mm$/$h and
$r_{k}=8$ mm$/$h, cf., Fig.~\ref{fig:finerain}. Of course, since fuel may dry
up between episodes of rain, in the case of intermittent rain, the result of
the model depends also on the temporal rain distribution, not only the total
accumulation -- just like fuel moisture in reality.

\label{sec:numerical}Because we want the model to support an arbitrarily long
time step, an adaptive exponential method was implemented. The method is exact
for long time step when the atmospheric variables and the rain intensity are
constant in time. Equations (\ref{eq:moisture-ode}) and (\ref{eq:rain-ode}%
)\ have the common form%
\begin{equation}
\frac{dm}{dt}=\frac{E-m}{T}. \label{eq:common}%
\end{equation}
For longer time steps, use the exact solution of (\ref{eq:common}) over the
time step interfal $\left[  t_{n},t_{n+1}\right]  $, with constant
coefficients taken as their values at $t_{n+1/2}=t_{n}+\Delta t/2$, which
gives $M_{n+1}=M_{n}+\left(  E_{n+1/2}-M_{n}\right)  \left(  1-e^{-\Delta
t/T_{n+1/2}}\right)  $, where $E_{n+1/2}$ and $T_{n+1/2}$ are computed at
$t_{n+1/2}$. This method is of second order and it is particularly useful when
the time step is comparable to or even larger than the time lag and the
coefficients $E$ and $T$ vary slowly, such as in the case of fast drying just
after the rain ends. For very short time steps, however, the rounding error in
the subtraction of the almost equal quantities in $1-e^{-\Delta t/T_{n+1/2}}$
will pollute the solution. Thus, for $\Delta t/T_{n+1/2}<\varepsilon=0.01$, we
replace the exponential by a truncated Taylor expansion, and use the second
order method $M_{n+1}=M_{n}+\left(  E_{n+1/2}-M_{n}\right)  \frac{\Delta
t}{T_{n+1/2}}\left(  1-\frac{1}{2}\frac{\Delta t}{T_{n+1/2}}\right)  .$

\begin{figure}[t]
{\centering
\begin{tabular}
[c]{cc}%
\hspace*{-0.5in} \includegraphics[height=3in]{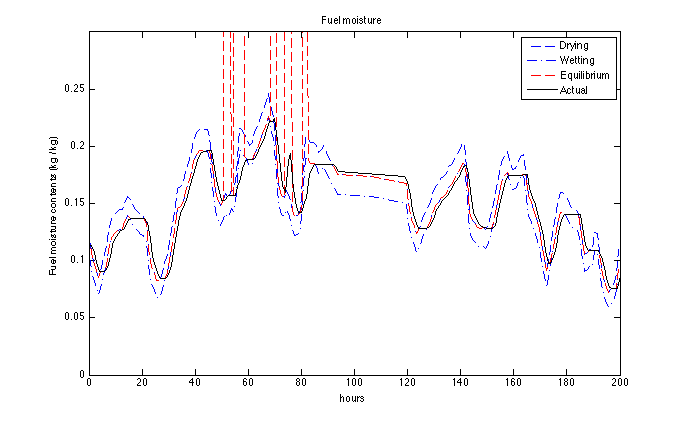}
\hspace*{-0.5in} &
\ \includegraphics[height=2.25in,trim=0 -100 0 0, clip]{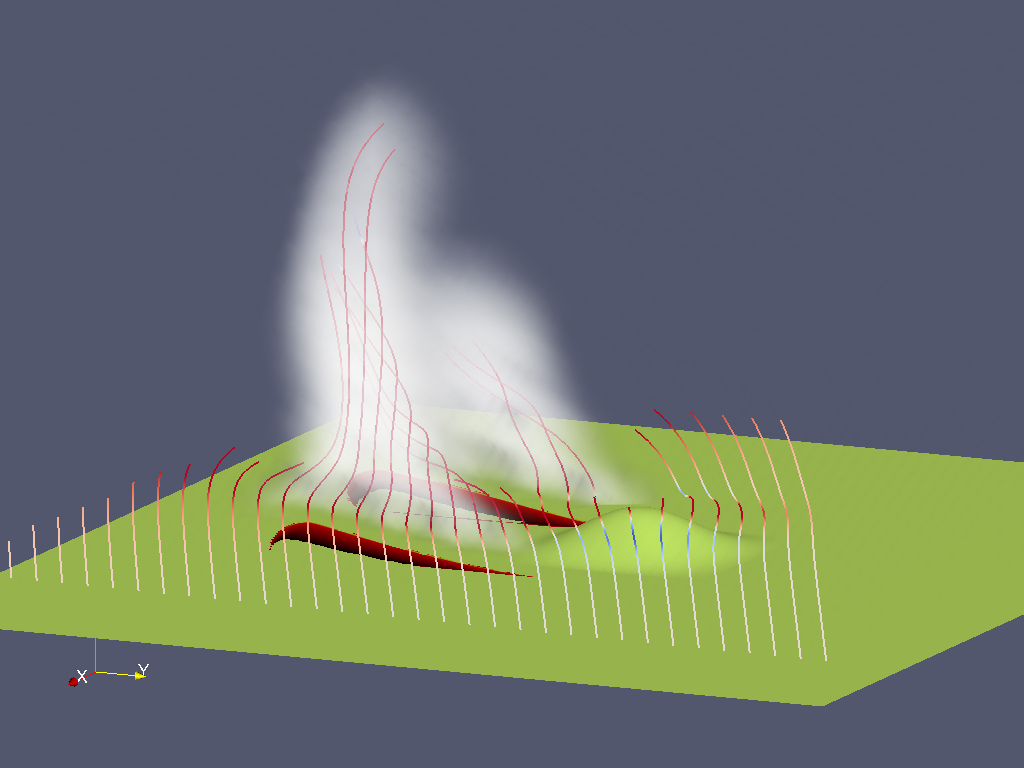}\\
(a) & (b)
\end{tabular}
} \caption{(a) Simulation of fuel moisture contents at a single point for fine
fuels. The drying and wetting equilibria are computed from the WRF state. The
fuel moisture contents does not change when it is between the two equilibria.
The red vertical lines correspond to periods of rain, where the equilibrium is
2.5 (above the range shown). The fuel moisture contents increases during rain
and exhibits diurnal variability. The flat part between 100 and 200 hours is due
to one day of missing data. (b) Advection of smoke by tracers injected
in WRF-Chem, in an ideal example.}%
\label{fig:2}%
\end{figure}

Because the time scale of the moisture changes (hours) is very different from
the time scales of the atmoshere (minutes(\ and fire (seconds), the moisture
model runs at a multiple of the WRF time step. WRF variables \texttt{P}
(pressure, Pa), \texttt{T2} (temperature at 2m, K), and \texttt{Q2} (water
vapor mixing ratio, kg/kg) at the beginning and the end of the moisture model
time step are averaged to obtain the values of $P$, $T$, and $Q$ used in the
timestep of the moisture model,%
\[
P_{n+1/2}=\frac{\mathtt{P}\left(  t_{n+1}\right)  +\mathtt{P}\left(
t_{n}\right)  }{2},\quad T_{n+1/2}=\frac{\mathtt{T2}\left(  t_{n+1}\right)
+\mathtt{T2}\left(  t_{n}\right)  }{2},\quad Q_{nn+1/2}=\frac{\mathtt{Q2}%
\left(  t_{n+1}\right)  +\mathtt{Q2}\left(  t_{n}\right)  }{2},
\]
This is done for 2nd order accuracy as well as for compatibility with the
computation of the rain intensity from the difference of the accumulated rain,%
\[
r_{n+1/2}=\frac{\mathtt{RAINC}(t_{n+1})+\mathtt{RAINNC}\left(  t_{n+1}\right)
-\mathtt{RAINC}(t_{n})-\mathtt{RAINNC}\left(  t_{n}\right)  }{\Delta t}.
\]
These values are then used to compute the equilibrium moisture contents and
the time lag, and one time step is performed, as described above. See
Fig.~\ref{fig:2}i(a) for a typical simulation result.

The fire model runs on a finer grid than the atmospheric model, typically
refined by a factor of $10$ or more. The moisture model consists of two steps.

In the first step, we compute the moisture content $m_{k}$ of the fuel
components on the nodes of the atmospheric grid on the Earth surface for
several reasons. (1) The WRF variables are known at the nodes of the
atmospheric grid and no interpolation is needed. (2) The atmospheric grid is
relatively coarse, so the added costs of storage of several surface moisture
fields $m_{k}$ and of the computation are not significant. (3) The computation
is done for each fuel component indivudually over the whole domain and it does
not depend on the actual fuel map. (4) The fire model does not need to run for
the computation of $m_{k}$.

In the second step, we interpolate the values of $m_{k}$ to the fire grid and
compute the weighted averages on the fire grid following using the actual fuel
map. There are several options how to run the fuel moisture model: (1) Turn on
both steps of the moisture model: compute the moisture fields $m_{k}$ and
interpolate on the fire grid, as the fire model runs. This option is intended
for the actual fire simulation. (2) Turn on the first step, computation of the
moisture fields $m_{k}$ only, for an extended run (many days) to evolve the
$m_{k}$ in response to the simulated weather, into a dynamic equilibrium. The
moisture fields are stored in WRF state in the output and the restart files.
(3) Run the first step, computation of the moisture fields $m_{k}$, as a
standalone executable from stored output files from standard WRF runs, adding
the moisture fields $m_{k}$ to the files. (4) Start the coupled
atmosphere-fire-moisture simulation from the moisture fields evolved over time
as above. (5) Run the fire model as a standalone exectutable, without feedback
on the atmosphere, with fuel moisture computed using the second step only from
the fields $m_{k}$ in WRF output files produced in advance.

\section{Coupling with WRF-Chem}

Coupling with WRF-Chem is implemented by inserting the smoke intensity in the
WRF-Chem arrays \texttt{emis\_ant} and \texttt{tracer} at the ground layer and
with the species index \texttt{p\_smoke}. WRF needs to be configured with the
\texttt{chem} option and built with \texttt{em\_fire} or \texttt{em\_real},
and the appropriate options \texttt{trace\_opt=1} and \texttt{chem\_opt=14}
set in the namelist. A sample visualization of an ideal run with smoke
transport is in Fig.~\ref{fig:2}(b). See the WRF SFIRE Users' Guide at
\url{http://www.openwfm.org/wiki/Users_guide} for further details and more
information as the code develops further.

Currently, the smoke inserted into WRF-Chem is simply proportional to the fire
heat flux. However, a more sophisticated scheme following the FEPS (Fire
Emission Production Simulator) is being implemented \cite{Anderson-2004-FEP}.
In this new method, the fire emission will be treated as the fluxes of carbon
dioxide (CO2), carbon monoxide (CO), methane (CH4) and particulate matter (PM
2.5) released into the atmosphere at the location of the fire. The total
emission for each of these species will be computed based on the fuel burn
rate (amount of fuel burnt per unit time), fuel load, and the smoldering
correction computed based from the wind speed and relative humidity. The PM
2.5 will be portioned into the the accumulation and nuclei mode and released
into the atmosphere as PM25J and PM25I respectively.

Aside from the obvious application to smoke dispersal, which is important in
practice, taking into account the composition of the fire emitted smoke will
allow for treating it as chemically and physically active. The species
released into the atmosphere by the fire will undergo chemical reactions
allowing for capturing the secondary aerosol effects. The particular mater
will interact with atmospheric radiation, and have a potential to serve as
condensation nuclei when suitable cloud microphysics is available. This may be
of great importance in the simulation of visible plumes and pyrocumulus
clouds, under development in \cite{Peace-2011-CSP}.

{\small
\bibliographystyle{ametsoc}
\bibliography{../../references/geo,../../references/other}
}

\end{document}